\documentclass[preprintnumbers, floatfix, letterpaper, aps, prd, epsfig, nofootinbib, twocolumn,
]{revtex4-1}
\usepackage{bm, graphicx, dcolumn, epstopdf, epsf, latexsym,mathbbol, amssymb, amsmath, color, slashed, mathrsfs, mathcomp, simplewick}
\usepackage[center]{subfigure}
\usepackage{multirow}
\usepackage{makecell}
\usepackage[colorlinks,linkcolor=blue,citecolor=blue,urlcolor=blue]{hyperref}
\begin{document}
\allowdisplaybreaks
%%%%%%%%%%%%%%%%%%%%%%%%
 \newcommand{\bq}{\begin{equation}}
 \newcommand{\eq}{\end{equation}}
 \newcommand{\bqn}{\begin{eqnarray}}
 \newcommand{\eqn}{\end{eqnarray}}
 \newcommand{\nb}{\nonumber}
 \newcommand{\lb}{\label}
 \newcommand{\f}{\frac}
 \newcommand{\p}{\partial}
%%%%%%%%%%%%%%%%%%%%%%%%%
\newcommand{\PRL}{Phys. Rev. Lett.}
\newcommand{\PLB}{Phys. Lett. B}
\newcommand{\PRD}{Phys. Rev. D}
\newcommand{\CQG}{Class. Quantum Grav.}
\newcommand{\JCAP}{J. Cosmol. Astropart. Phys.}
\newcommand{\JHEP}{J. High. Energy. Phys.}
\newcommand{\NPB}{Nucl. Phys. B}
\newcommand{\Doi}{https://doi.org}
\newcommand{\arXiv}{https://arxiv.org/abs}
\newcommand{\red}{\textcolor{red}}
 %%%%%%%%%%%%%%%%%%%%%%%%
\title{Inflationary perturbation spectra at next-to-leading slow-roll order in effective field theory of inflation}

\author{Guang-Hua Ding${}^{a}$}
\email{dingguanghua@zjut.edu.cn}

\author{Jin Qiao ${}^{a}$}
\email{qiaojin@zjut.edu.cn}

\author{Qiang Wu${}^{a}$}
\email{wuq@zjut.edu.cn}

\author{Tao Zhu${}^{a}$}
\email{zhut05@zjut.edu.cn; Corresponding author}

\author{Anzhong Wang${}^{a, b}$}
\email{anzhong\_wang@baylor.edu}

\affiliation{${}^{a}$ Institute for Theoretical Physics $\&$ Cosmology, Zhejiang University of Technology, Hangzhou, 310032, China \\
${}^{b}$ GCAP-CASPER, Physics Department, Baylor University, Waco, TX 76798-7316, USA
}

\date{\today}

\begin{abstract}

The effective field theory (EFT) of inflation provides an essential picture to explore the effects of the unknown high energy physics in the single scalar field inflation models. For a generic EFT of inflation, possible high energy corrections to simple slow-roll inflation can modify both the propagating speed and dispersion relations of the cosmological scalar and tensor perturbations. With the arrival of the era of precision cosmology, it is expected that these high energy corrections become more important and have to be taken into account in the analysis with future precise observational data. In this paper we study the observational predictions of the EFT of inflation by using the third-order uniform asymptotic approximation method. We calculate explicitly the primordial power spectra, spectral indices, running of the spectral indices for both scalar and tensor perturbations, and the ratio between tensor and scalar spectra. These expressions are all written in terms of the Hubble flow parameters and the flow of four new slow-roll parameters and expanded up to the next-to-leading order in the slow-roll expansions so they represent the most accurate results obtained so far in the literature. The flow of the four new slow-roll parameters, which arise from the four new operators introduced in the action of the EFT of inflation, can affect the primordial perturbation spectra at the leading-order and the corresponding spectral indices at the next-to-leading order.
%The running of the indices are keeping the same with that in the standard slow-roll inflation up to the next-to-leading order.

\end{abstract}
\maketitle

%%%%%%%%%%%%%%%%%%%%%%%%%%%%%%%
%%%%%%%%%%%%%%%%%%%%%%%%%%%%%%%
\section{Introduction}
\renewcommand{\theequation}{1.\arabic{equation}}\setcounter{equation}{0}
%%%%%%%%%%%%%%%%%%%%%%%%%%%%%%%
%%%%%%%%%%%%%%%%%%%%%%%%%%%%%%%

Inflation is the leading paradigm for the hot Big Bang origin of the Universe \cite{guth_inflationary_1981, starobinsky_new_1980, sato_firstorder_1981} (see Ref.~\cite{baumann_tasi_2009} for an updated review). The most remarkable features of the inflationary scenario is that it provides an elegant mechanism for generating structures in the Universe and the spectrum of cosmic microwave background (CMB) anisotropies, which are fully consistent with the cosmological observations with a spectacular precision \cite{komatsu_sevenyear_2011, planckcollaboration_planck_2018, planck_collaboration_planck_2015-4, planck_collaboration_planck_2014-1}.  In general, there are a lot of approaches to realize inflation that originate from very different background physics. The recent precise observational data set supports the key predictions of the standard single-field inflationary models. In this model, a single scalar degrees of freedom with a self-interacting potential produces a slow-roll phase during which the energy density of the matter field remains nearly constant and the space-time behaves like a quasi-de Sitter spacetime.

Instead of constructing inflation models with scalar degrees of freedom from different fundamental theories, the EFT provides a general framework for describing the most generic single scalar field theory and the associated fluctuations on a quasi de-Sitter background \cite{cheung_effective_2008, weinberg_effective_2008}. This framework is based on the { decomposition of space} and time with a preferred time foliation defined by the scalar field, which provides a clock that breaks time but preserves spatial diffeomorphism invariance. Similar idea has been used in the construction of the Horava-Lifshitz (HL) theory of quantum gravity \cite{horava_quantum_2009, horava_general_2010, zhu_symmetry_2011, zhu_general_2012, lin_postnewtonian_2014}, in which the symmetry of the theory is broken from the general covariance down to the foliation-preserving diffeomorphisms. With this symmetry, the action of the theory has to be constructed only in terms of three dimensional spatial diffeomorphism invariants. This allows us to characterize all the possible high energy corrections to simple slow-roll inflation and their impacts on the primordial perturbation spectra. Such considerations have attracted a lot of attention and the observational effects of high-order operators on inflationary perturbation spectra have been extensively studied both in the framework of the EFT of inflation and the inflation in HL theory \cite{huang_primordial_2013, zhu_inflation_2013, zhu_effects_2013, wang_polarizing_2012} (see \cite{wang_horava_2017} for an updated review). With such thoughts, an extension of the EFT of inflation by adding high dimension operators has also been constructed and the corresponding primordial perturbation spectra have also been explored in refs. \cite{ashoorioon_extended_2018, ashoorioon_nonunitary_2018, qiao_inflationary_2018}.

In the EFT of inflation, possible high energy corrections to simple slow-roll inflation can affect the propagations of both the cosmological scalar and tensor perturbations in two aspects.  One is that it may lead to a time-dependent propagating sound speeds associated with the scalar and tensor perturbation modes, and another is that it could modify the conventional linear dispersion relation of the perturbation modes to nonlinear ones.  Both the time-dependent  propagating sound speeds and the modified dispersion relation can make important corrections in the primordial scalar and tensor perturbation spectra. In most of previous works, however, in order to estimate the spectra, the time-dependent speed has been assumed to be constant and the operators which produces nonlinear dispersion relations are ignored \cite{cheung_effective_2008, cheung_consistency_2008, naskar_inflation_2017, choudhury_cmb_2017}. One of the main reasons for such treatment is that the considerations of the time variation of the sound speed and the nonlinear dispersion relations make it very difficult to calculate the corresponding power spectra and spectral indices.

However, to match with the accuracy of the current and forthcoming observations, as pointed out in \cite{wu_primordial_2018, martin_kinflationary_2013, martin_encyclopaedia_2014, martin_shortcomings_2016}, consideration of the contributions from the time variation of the sound speed and the modified dispersion relation are highly demanded. To achieve this goal, one has to calculate the primordial perturbation spectra up to the second-order in the slow-roll expansions. Recently, we have developed a powerful method, the uniform asymptotic approximation method \cite{zhu_constructing_2014, zhu_inflationary_2014, zhu_quantum_2014}, to calculate precisely the primordial perturbation spectra for a lot of inflation models. The robustness of this method has been verified for calculating primordial spectra in k-inflation \cite{zhu_power_2014, martin_kinflationary_2013,ringeval_diracborninfeld_2010}, and inflation with nonlinear dispersion relations \cite{zhu_inflationary_2014, zhu_quantum_2014, zhu_highorder_2016}, inflation in loop quantum cosmology \cite{zhu_scalar_2015, zhu_detecting_2015, zhu_inflationary_2016}, and inflation with Gauss-Bonnet corrections \cite{wu_primordial_2018}  (For an alternative approach by using Green’s function method, see \cite{stewart_density_2001, wei_secondorder_2004} and its recent application to the scalar-tensor theories in different frames \cite{karam_framedependence_2017}.). We note here that this method was first applied to inflationary cosmology in the framework of general relativity (GR) in \cite{habib_inflationary_2002, habib_inflationary_2005, habib_characterizing_2004}, and then we have developed it, so it can be applied to more general cases and high accuracy \cite{zhu_inflationary_2014,zhu_highorder_2016}. The accurate results of some inflationary models derived from this method have also been used to the study of the adiabatic regularisation of the primordial power spectra \cite{alinea_adiabatic_2015, alinea_adiabatic_2016}. It is {worth noting} that this approximation {has also been applied to} study the parametric resonance during reheating \cite{zhu_field_2018a} and the quantization conditions in quantum mechanics \cite{zhu_langer_2019a}. The main purpose of the present paper is to use this powerful method to derive the inflationary observables at the second-order in slow-roll inflation in the framework of the EFT of inflation. With the general expressions of power spectra and spectral indices obtained in \cite{zhu_power_2014, zhu_quantum_2014} in the uniform asymptotic approximation, we calculate explicitly these quantities for both scalar and tensor perturbations of EFT of inflation. Then, ratio between the tensor and scalar spectra is also calculated up to the second-order in the slow-roll expansions. These observables represent a significant improvement over the previous results obtained so far in the literature.

We organize the rest of the paper as follows. In Sec.~II, we present a brief review of the EFT of inflation and the corresponding equations of motion for both cosmological scalar and tensor perturbations. In Sec.~III, we give the most general formulas of the primordial perturbation spectra in the uniform asymptotic approximation. Then, in Sec. IV, with these general expressions we calculate explicitly the power spectra, spectral indices, and running of the spectral indices of both scalar and tensor perturbations of the slow-roll inflation in the EFT of inflation. Our main conclusions and  discussions are presented  in Sec. V.

\section{The effective field theory of inflation}
\renewcommand{\theequation}{2.\arabic{equation}}\setcounter{equation}{0}

In this section, we present a brief introduction of the EFT of inflation \cite{cheung_effective_2008, weinberg_effective_2008}. In general, the EFT provides a framework for describing the most generic single scalar field theory on a quasi de-Sitter background. With this framework, it is shown that the action of the EFT of inflation around a flat Friedmann-Lema\'itre-Robertson-Walker (FLRW) background reads
\begin{widetext}
\bqn\lb{Seft}
S_{\rm eft}&=&M_{\rm Pl}^2 \int d^4 x \sqrt{-g} \Bigg\{\frac{R}{2}+\dot H g^{00} - (3H^2+\dot H) + \frac{M_2^4}{2 M_{\rm Pl}^2}(g^{00}+1)^2 -\frac{\bar M_1^3}{2 M_{\rm Pl}^2} (g^{00}+1)\delta K^\mu_\mu   \nb\\
&& ~~~~~~~~~~~~~~~~~~~~~~  - \frac{\bar M_2^2}{2 M_{\rm Pl}^2} (\delta K_\mu^\mu)^2- \frac{\bar M_3^2}{2M_{\rm Pl}^2} \delta K^\mu_\nu \delta K^\nu_\mu\Bigg\},
\eqn
\end{widetext}
where 
$$K_{\mu\nu} = (g_{\mu  \sigma} + n_\mu n_\sigma) \nabla^\sigma n_\nu$$
 denotes the extrinsic curvature at constant time hypersurfaces with $n_\nu$ being the unit normal vector. $\delta K_{\mu\nu}$ denotes the perturbation of $K_{\mu\nu}$ about the flat FLRW background. For the calculation of the power spectra for both the scalar and tensor perturbation later, we only consider operators up to second order fluctuation in metric, which is sufficient to extract the information of two-point function from scalar and tensor modes. Therefore, we truncate the EFT action to (\ref{Seft}) in the above by cutting terms, for example, with $n >2$ in $\hat{\mathcal{O}}_{(i)}=\sum\limits_{n=2}^{\infty}\frac{M_n^4(t)}{n!}(1+g^{00})^n$.
 
 We would like to mention that, in writing the above action, one requires the perturbation in inflaton field $\phi$ to be zero, i.e., $\delta \phi=0$. This is achieved by considering the following linear transformations under time diffeomorphism,
\bqn
\tilde t = t+\xi^0(t,x^i), \;\;  \delta \tilde \phi = \delta  \phi + \xi^0(t, x^i) \dot \phi_0(t),
\eqn
which leads to a particular gauge (unitary gauge) with $\xi^0(t,x^i)= - \delta\phi/\dot \phi_0$ where there is no inflaton perturbation. Obviously the action (\ref{Seft})  that respects the unitary gauge has no time diffeomorphism invariance. In order to restore the time diffeomorphism invariance, one can introduce a Goldstone mode $\pi(t,x^i)$ and require it transforms as $\pi(t,x^i) \to \pi(t,x^i)- \xi^0(t,x^i)$. By introducing the Goldstone mode, the perturbation of inflaton field is not required to be zero and it relates to $\pi(t,x^i)$ via $\delta \phi = \dot \phi_0 \pi$.

It is {worth noting that} the standard single scalar slow-roll inflation can be simply recovered by setting $M_2, \; \bar M_1, \bar M_2, \bar M_3$ to zero. As mentioned in \cite{cheung_effective_2008}, {the terms with $M_2, \; \bar M_1, \bar M_2, \bar M_3$ encodes the possible effects of high energy physics on the simple slow-roll model of inflation under the current accurate observation.} As we will see later, these new terms can affect both the propagating speed and dispersion relations of the cosmological scalar and tensor perturbations. Therefore, studying their effects on the primordial perturbation spectra can provide a way to probe the possible high energy effects in the observational data.

\subsection{Scalar perturbations}

As mentioned above, the Goldstone mode $\pi(t,x^i)$ can be introduced to restore the time diffeomorphism invariance of action and it also describes the scalar perturbations around the flat FLRW background. Thus, in order to study the scalar perturbations, one can transform the action in (\ref{Seft}) in unitary gauge to $\pi$-gauge by evaluating the action explicitly for $\pi$. 
%{In order to get the second order of $r$, we only need to calculate the first-order of the power spectrum, so we do not consider the term of $\delta g^{00}$ or $\delta g^{0i}$ and only consider the effect of $g^{00}$ to the second-order.} 
Proceeding this procedure \cite{cheung_effective_2008}, one obtains
\bqn\lb{pi_action}
S^\pi_{\rm eft} &=& \int d^4x a^3 \Big[A_0 (\dot \pi^2 +  3 H^2 \epsilon_1 \pi^2 )+ A_1 \frac{(\partial_i \pi)^2}{a^2} \nb\\
&&~~~~~\;\; + A_2  \frac{(\partial^2\pi)^2}{a^4} \Big],
\eqn
where
%\bqn
%\mathcal{L}_{\rm eff} &=& M^{2}_{\rm Pl}\dot{H}(\partial_{\mu}\pi)^{2}+2M^{4}_{2}\dot{\pi}^{2}-\bar{M}^{3}_{1}H \left(3\dot{\pi}^{2}-\frac{(\partial_{i}\pi)^{2}}{2a^{2}}\right)  \nb\\
%&&-\frac{\bar{M}^{2}_{2}}{2}\left(9H^{2}\dot{\pi}^{2}-3H^{2}\frac{(\partial_{i}\pi)^{2}}{a^{2}}+\frac{(\partial^{2}_{i}\pi)^{2}}{a^{4}}\right)  \nb\\
%&&-\frac{\bar{M}^{2}_{3}}{2} \left(3H^{2}\dot{\pi}^{2}-H^{2}\frac{(\partial_{i}\pi)^{2}}{a^{2}}+\frac{(\partial^{2}_{j}\pi)^{2}}{a^{4}}\right).
%\eqn
%Variation the above action $S^\pi_{\rm eft}$ with respect to $\pi$ yields the equation of motion for the Goldstone mode $\pi$,
%\bqn
%A_0 \ddot \pi_k + (B_0+3 H A_0) \dot \pi_k +\left(D_0 \frac{k^2}{a^2}+E_0 \frac{k^4}{a^4}\right)\pi_k=0,\nb\\
%\eqn
%where
\bqn
A_0 &=& -M^{2}_{\rm Pl}\dot{H}+2M^{4}_{2}-3\bar{M}^{3}_{1}H-\frac{9}{2}H^{2}\bar{M}^{2}_{2}\nb\\
&&-\frac{3}{2}H^{2}\bar{M}^{2}_{3}, \\
A_1 &=&  M^{2}_{\rm Pl}\dot{H}+ \frac{1}{2}\bar{M}^{3}_{1}H + \frac{3}{2}H^{2}\bar{M}^{2}_{2}+\frac{1}{2}\bar{M}^{2}_{3}H^{2},\\
A_2 &=& - \frac{1}{2} \left(\bar M_2^2 +\bar M_3^2 \right).
%B_0 &=&-6\bar{M}^{3}_{1}\dot{H}-18\dot{H}H\bar{M}^{2}_{2}-6\dot{H}H\bar{M}^{2}_{3},\\
%C_0 &=& -2M^{2}_{Pl}\dot{H}+4M^{4}_{2}-6\bar{M}^{3}_{1}H-9H^{2}\bar{M}^{2}_{2},\\
%D_0 &=& -2M^{2}_{Pl}\dot{H}-\bar{M}^{3}_{1}H-3H^{2}\bar{M}^{2}_{2}-\bar{M}^{2}_{3}H^{2},\\
%E_0 &=&\bar{M}^{2}_{2}+\bar{M}^{2}_{3}.
\eqn
Here two remarks about the above action are in order. First, the above action for scalar perturbations is valid in the weak decoupling limit. In this way, we ignore most of the terms arising from mixing between gravity and Goldstone Boson, except the leading-order mixing term $M_{\rm Pl}^2 \dot H \dot \pi \delta g^{00}$. This term gives rise to a mass term $3 a^3 H^2 \epsilon_1 \pi^2$ in the action for the scalar perturbations. Second, in order to match the precision of the current and future observations, both the scalar spectral index $n_s$ and tensor-to-scalar ratio $r$ are required be accurate up to the second-order in the slow-roll expansion. In this way, in order to simplify our calculations, we only interest in terms in the action which contribute to $n_s$ and $r$ at the first two orders in the slow-roll expansion. Other terms which lead to contributions beyond this order are all neglected in the above action.

Changing the variable $\pi$ to $u = z \pi $ we cast the action into the form
\bqn
S^\pi_{\rm eft} &=& \int d\eta d^3x \Big[ u'^2 + \left(\frac{z''}{z} + 3 a^2 H^2 \epsilon_1 \right)u^2 + \frac{A_1}{A_0} (\partial_i u)^2 \nb\\
&&~~~~~~~~  + \frac{A_2}{A_0}  \frac{(\partial^2 u)^2}{a^2} \Big],
\eqn
where a prime denotes the derivative with respect to the conformal time $\eta$ and
\bqn
z = a \sqrt{A_0}.
\eqn
Then variation of the action with respect to the Fourier modes $u_k(\eta)$ of $u$ leads to the equation of motion,
\bqn \lb{1_scalar}
u''_k + \left( \omega_k^2(\eta)-\frac{z''}{z}-3 a^2 H^2 \epsilon _1\right)u_k=0,
\eqn
where
\bqn
&&c_s^2 \equiv - \frac{A_1}{A_0},
\eqn
and
\bqn\lb{nonlinear}
\omega_k^2(\eta) = c_s^2 k^2 \left(1- \frac{A_2}{A_0 c_s^2}  \frac{k^2}{a^2} \right).
\eqn
It is easy to see that in the EFT of inflation, both the effective sound speed and the nonlinear terms in the dispersion relation receive contributions from the operators with $M_2, \bar M_1, \bar M_2, \bar M_3$. If one ignores $\bar M_2$ and $\bar M_3$ terms, the nonlinear dispersion relation reduces to the usual linear one. In refs. \cite{cheung_consistency_2008, cheung_effective_2008, choudhury_cmb_2017, naskar_inflation_2017}, the nonlinear term ($A_2$ term in (\ref{nonlinear}))  has been dropped and the time-dependent sound speed $c_s$ has been assumed to be constant during inflation. In this paper, we are going to consider effects of both the nonlinear term and the time variation of the sound speed in the primordial power spectra.

\subsection{Tensor perturbations}

For tensor perturbations, the perturbed spacetime is set as
\bqn
g_{ij}=a^2 (\delta_{ij}+h_{ij}),
\eqn
where $h_{ij}$ represents the transverse and traceless tensor perturbations, which satisfies
\bqn
h^{i}_{i}=0=\partial^i h_{ij}.
\eqn
Then expanding the total action $S_{\rm eft}$ up to the second-order gives
\bqn
S^{2}_{h} = \frac{M_{\rm Pl}^2}{8} \int dt d^3x a^3 \left(c_t^{-2}\partial_t h_{ij} \partial_t h^{ij}-a^{-2} \partial_k h_{ij} \partial^k h^{ij}\right),\nb\\
\eqn
where the effective sound speed $c_t$ for tensor perturbations is given by
\bqn
\lb{ct}
c_t^2 = \left(1-\frac{\bar M_3^2}{M_{\rm Pl}^2}\right)^{-1}.
\eqn
In order to avoid superluminal propagation for tensor modes, one has to require
\bqn
\bar M_3^2 <0.
\eqn
Variation of the action with respect to $h_{ij}$ leads to the equation of motion
\bqn
\lb{1_tensor}
\frac{d^2\mu_k^{(t)}(\eta)}{d\eta^2} + \left(c_t^2 k^2- \frac{a''}{a}\right) \mu_k^{(t)}(\eta)=0,
\eqn
where $d\eta =dt/a$ is the conformal time and $\mu_k^{(t)}(\eta)=a M_{\rm Pl} h_k/\sqrt{2}$ with $h_k$ denoting the Fourier modes of the two helicities of the tensor perturbations.

It is obvious to observe that the new operators introduced in the action (\ref{Seft}) of EFT of inflation can affect sound speed of the tensor modes. As pointed out in \cite{one, two}, this effective sound speed can be redefined to unity by a disformal transformation. Under this disformal transformation, the new effects in the sound speed shift to the effective time-dependent mass term $a''/a \to \tilde a''/\tilde a$ in (\ref{1_tensor}) such that the primordial tensor spectrum keeps invariant. With this property, one can either do the calculation with $c_t \neq 1$ or with $c_t=1$ by using the disformal transformation. In this paper, we adopt the former case.

\subsection{Slow-roll parameters}

In order to consider the slow-roll inflation, we first need to impose the following slow-roll conditions,
\bqn
\left| \frac{\dot H}{H^2} \right|,\;\;  \left|\frac{\ddot H}{H \dot H}\right| \ll 1.
\eqn
With these conditions, it is convenient to introduce a set of the slow-roll parameters, the Hubble flow parameters, which are defined by
\bqn\lb{ff}
\epsilon_1 \equiv - \frac{\dot H}{H^2},\;\; \epsilon_{n+1} \equiv \frac{d \ln \epsilon_n}{d \ln a}.
\eqn
In the EFT of inflation, in general one also requires all the coefficients $(M_2^4, \bar M_1^3, \bar M_2^2, \bar M_3^2)$ of the new operators in the action (\ref{Seft}) satisfy the approximation conditions:
\bqn
\left|\frac{M_2^4}{M_{\rm Pl}^2 H^2 \epsilon_1}\right|, \left|\frac{\bar M_1^3}{M_{\rm Pl}^2 H \epsilon_1}\right|,  \left|\frac{\bar M_2^2}{M_{\rm Pl}^2  \epsilon_1}\right|,  \left|\frac{\bar M_3^2}{M_{\rm Pl}^2  \epsilon_1}\right| \ll 1.\nb\\
\eqn
In the slow-roll approximation, we also need all the above quantities are slow-varying. With this assumption, we can introduce four new sets of slow-roll parameters, which are defined by
\bqn
\delta_1 \equiv  \frac{M_2^4}{M_{\rm Pl}^2 H^2 \epsilon_1}, \;\; \delta_{n+1} = \frac{d \ln \delta_{n}}{d \ln a},\\
\sigma_1 \equiv  \frac{\bar M_1^3}{M_{\rm Pl}^2 H \epsilon_1}, \;\; \sigma_{n+1} = \frac{d \ln \sigma_{n}}{d \ln a},\\
\zeta_1 \equiv  \frac{\bar M_2^2}{M_{\rm Pl}^2  \epsilon_1}, \;\; \zeta_{n+1} = \frac{d \ln \zeta_{n}}{d \ln a},\\
\kappa_1 \equiv  \frac{\bar M_3^2}{M_{\rm Pl}^2  \epsilon_1}, \;\; \kappa_{n+1} = \frac{d \ln \kappa_{n}}{d \ln a}.
\eqn

Then using these slow-roll parameters, up to the second-order, the effective sound speed for scalar perturbations and the modified dispersion relation $\omega_k^2(\eta)$ can be expressed as
\bqn\lb{fff}
\lb{cs} c_s^2 &=& - \frac{3 \zeta _1+\kappa _1+\sigma _1-2}{4 \delta _1-9 \zeta _1-3 \kappa _1-6 \sigma _1+2}\nb\\
&\simeq &1+\frac{1}{4} \left(-4 \delta _1+6 \zeta _1+2 \kappa _1+5 \sigma _1\right)\nb\\
&&-6 \delta _1 \zeta _1-2 \delta _1 \kappa _1-\frac{17 \delta _1 \sigma _1}{4}+\frac{3 \delta _1^2}{2}\nb\\
&&+\frac{15 \zeta _1 \kappa _1}{4}+\frac{33 \zeta _1 \sigma
   _1}{4}+\frac{45 \zeta _1^2}{8}\nb\\
   &&+\frac{11 \kappa _1 \sigma _1}{4}+\frac{5 \kappa _1^2}{8}+\frac{95 \sigma _1^2}{32},\\
\nb\\
\omega_k^2 (\eta) &\simeq& c_s^2 k^2 + b(\eta) k^4 \eta^2,
\eqn
where $b(\eta)$ is a small and slow-varying quantity which reads
\bqn
b(\eta) & \simeq & \frac{\zeta_1+\kappa_1 }{2}-\zeta _1 \epsilon _1-\kappa _1 \epsilon _1-\delta _1 \zeta _1-\delta _1 \kappa _1\nb\\
&+&\frac{3 \zeta _1 \kappa _1}{2}+\frac{3 \zeta _1 \sigma _1}{2}+\frac{9 \zeta _1^2}{4}+\frac{3 \kappa _1 \sigma _1}{2}-\frac{3\kappa _1^2}{4}.\nb\\
\eqn

\section{Scalar and Tensor Perturbation Spectra in the uniform asymptotic approximation}
\renewcommand{\theequation}{3.\arabic{equation}}\setcounter{equation}{0}
%%%%%%%%%%%%%%%%%%%%%%%%%%%%%%%
%%%%%%%%%%%%%%%%%%%%%%%%%%%%%%%

\subsection{General formulas of primordial spectra in the uniform asymptotic approximation}

In this subsection, we present a very brief introduction of the general formulas of primordial perturbations with a slow-varying sound speed and parameter $b(\eta)$.

In the uniform asymptotic approximation, we first write Eqs.(\ref{1_scalar}) and (\ref{1_tensor}) in the standard form \cite{zhu_inflationary_2014, olver_asymptotics_1997}
\bqn
\frac{d^2 \mu(y)}{dy^2}=\{\lambda^2 \hat g(y)+q(y)\}\mu(y),
\eqn
where we introduce a new variable $y=-  k \eta$, $\mu(y)=\mu_{\mathcal{R}}(y)\;\text{and}\;\mu_h(y)$ corresponding to scalar and tensor perturbations respectively, and
\bqn
\lambda^2 \hat g(y)+q(y)=\frac{\nu^2(\eta)-1/4}{y^2}-c^2(\eta) - b^2(\eta) y^2.
\eqn
Here $c(\eta) = c_{s ,t}(\eta)$ is the effective sound speed for scalar and tensor perturbation modes respectively.  For scalar perturbation, we have \cite{choudhury_cmb_2017, cheung_consistency_2008}
\bqn\lb{zs}
\nu_s^2(\eta) = \eta^2 \frac{z''(\eta)}{z(\eta)} + \frac{1}{4}+3 a^2 H^2 \epsilon _1,
\eqn
and for tensor modes,
\bqn\lb{zt}
\nu_t^2(\eta) = \eta^2 \frac{a''(\eta)}{a(\eta)} + \frac{1}{4} \;\; {\rm and}\;\; b(\eta)=0.
\eqn
Note that in the above equation, $\lambda$ is supposed to be a large parameter and used to label the different orders of the approximation. As we will see later in the general formulas of power spectrum in Eq.~(\ref{formula_pw}), $\lambda$ with different power in the terms in square bracket denote different approximate orders in the uniform asymptotic approximation. In the finial calculation we can set $\lambda=1$ for simplification. Now in order to construct the approximate solutions of the above equation
by using the uniform asymptotic approximation, one needs to choose \cite{zhu_inflationary_2014}
\bqn
q(y)=- \frac{1}{4 y^2},
\eqn
to ensure the convergence of the errors of the approximate solutions. Then,  we have
\bqn
\lambda^2 \hat g(y)=\frac{\nu^2(\eta)}{y^2}-c^2(\eta) - b(\eta) y^2.
\eqn
In order to get a healthy UV limit, for scalar perturbation we have to impose the condition $b(\eta) >0$.

Considering that $b(\eta)$ is a small quantity, it is obvious to see that the function $\lambda^2 \hat g(y)$ has a single turning point,  which can be expressed as
\bqn
y_0^2(\bar \eta_0) &=&\frac{-c_0^2(\bar \eta_0) + \sqrt{\bar c_0^4(\eta_0) + 4 b(\bar \eta_0) \bar \nu_0^2(\bar \eta_0)}}{2 b(\bar \eta_0)}.
\eqn
Then following \cite{zhu_power_2014}, the general formula of the power spectrum reads
\bqn\lb{formula_pw}
\Delta^2(k) &\equiv& \frac{k^3H^2}{4\pi^2} \left|\frac{u(y)}{z(\eta)}\right|^2_{y\to 0^+}\nb\\
&=&\frac{k^2}{8\pi^2}\frac{-k \eta}{z^2(\eta)\nu(\eta)}\exp{\left(2 \lambda  \int_y^{\bar y_0} \sqrt{\hat g(y')}dy' \right)}\nb\\
&&\times \left[1+\frac{\mathscr{H}(+\infty)}{\lambda}+\frac{\mathscr{H}^2(+\infty)}{2 \lambda^2}+\mathcal{O}\left(\frac{1}{\lambda^3}\right)\right].\nb\\
\eqn
We would like to mention that, in the square bracket of the above equation, the parameter $\lambda$ with different powers denote different approximate orders in the uniform asymptotic approximation, which shows clearly the above formula is at the third-order approximation. In this formulas, the integral of $\sqrt{g}$ is given by Eqs.~(\ref{intg}) with $I_0$ and $I_1$ being given by Eqs.~(\ref{I0}) and (\ref{I1}) and the error control function is given by (\ref{error}) with $\mathscr{H}_0$ and $\mathscr{H}_1$ being given by Eqs.~(\ref{H0}) and (\ref{H1}).

Now we turn to consider the corresponding spectral indices. In order to do this, we  first specify the $k$-dependence of $\bar\nu_0(\eta_0)$, $\bar \nu_1(\eta_0)$ through  $\eta_0 = \eta_0(k)$. From the relation $-k\eta_0= \bar y_0$, we have
\bqn
\frac{d\ln(-\eta_0)}{d\ln k} & =& -1 +\frac{d \ln \bar y_0}{d\ln (-\eta_0)}\frac{d \ln(-\eta_0)}{d \ln k},
\eqn
which leads to
\bqn
\frac{d\ln(-\eta_0)}{d\ln k} & \simeq &-1 -\frac{d \ln \bar y_0}{d\ln (-\eta_0)}.
\eqn
Then using this relation, the spectral index is given by
\bqn\lb{index}
n_{s}-1, \; n_t &\equiv& \frac{d\ln \Delta^2_{s,t}}{d \ln k} \nb\\
&\simeq &3-2\bar{\nu}_0-\frac{2\bar{b}_1\bar{\nu}_0}{3\bar{c}_0^4}+\frac{2\bar{c}_1\bar{\nu}_0}{\bar{c}_0}+\frac{2\bar{b}_1\bar{\nu}_0^3}{3\bar{c}_0^4}\nb\\
&&-2\bar{\nu}_1\ln2+\frac{\bar{\nu}_1}{6\bar{\nu}_0^2}.
\eqn
Similarly, we find that the running of the spectral index $\alpha \equiv dn/d\ln k$ is given by
\bqn \lb{running}
\alpha_{s,\; t}(k) &\simeq&2\bar{\nu}_1+\frac{2\bar{b}_2\bar{\nu}_0}{3\bar{c}_0^4}-\frac{2\bar{c}_2\bar{\nu}_0}{\bar{c}_0}-\frac{2\bar{b}_2\bar{\nu}_0^3}{3\bar{c}_0^4}\nb\\
&&+2\bar{\nu}_2\ln2-\frac{\bar{\nu}_2}{6\bar{\nu}_0^2}.
\eqn

In the above, we present all the formulas (Eqs.(\ref{formula_pw}), (\ref{index}), and (\ref{running})) that can be directly used to calculate the primordial perturbation spectra from different inflation models. Note
 that these formulas are easy to use because they only depend on the quantities $H(\eta)$, ($c_0, c_1, c_2$), ($\nu_0, \nu_1, \nu_2$) evaluated at the turning point. These quantities can be easily calculated from Eqs.~(\ref{cs}, \ref{zs}) for scalar perturbations and Eqs.~(\ref{ct}, \ref{zt}) for tensor perturbations. In the following subsections, we apply these formulas to calculate the slow-roll power spectra for both scalar and tensor perturbations in the EFT of inflation.

\subsection{Scalar Spectrum}

We first consider the scalar perturbations. As we already pointed out in the introduction, in order to match the accuracy of forthcoming observations, we need to calculate the spectral indices up to the next-to-leading order (second-order) in the expansions of the slow-roll approximation. For this purpose, we only need to consider the quantities $(\bar c_s, \; c_{\mathcal{R} 1})$ and $(\nu_{\mathcal{R} 0}, \; \nu_{\mathcal{R} 1}, \; \nu_{\mathcal{R} 2})$ up to the second-order in the slow-roll expansions.

For the slow-varying sound speed $c_{\mathcal{R}}$, from Eq.(\ref{cs}) we find
\bqn
\bar c_{\mathcal{R}0}& \simeq &1-\bar{\delta}_1+\frac{3\bar{\zeta}_1}{2}+\frac{\bar{\kappa}_1}{2}+\frac{5\bar{\sigma}_1}{4}-6\bar{\delta}_1\bar{\zeta}_1-2\bar{\delta}_1\bar{\kappa}_1\nb\\
&&-\frac{17\bar{\delta}_1\bar{\sigma}_1}{4}+\frac{3\bar{\delta}_1^2}{2}+\frac{15\bar{\zeta}_1\bar{\kappa}_1}{4}+\frac{33\bar{\zeta}_1\bar{\sigma}_1}{4}\nb\\
&&+\frac{45\bar{\zeta}_1^2}{8}+\frac{11\bar{\kappa}_1\bar{\sigma}_1}{4}+\frac{5\bar{\kappa}_1^2}{8}+\frac{95\bar{\sigma}_1^2}{32}.
\eqn
For $\nu_{\mathcal{R}}$, from Eqs.(\ref{nu}) and (\ref{zs}) we find
%Then one gets
\bqn
\bar \nu_{\mathcal{R} 0}& \simeq&\frac{3}{2}+\bar{\epsilon} _1+\frac{\bar{\epsilon} _2}{2}+\bar{\epsilon} _1^2+\frac{11 \bar{\epsilon} _1 \bar{\epsilon}
   _2}{6}+\frac{\bar{\epsilon} _2 \bar{\epsilon} _3}{6}\nb\\
&&+\bar{\delta} _1 \bar{\delta} _2-\frac{9 \bar{\zeta} _1 \bar{\zeta} _2}{4}-\frac{3 \bar{\kappa} _1 \bar{\kappa} _2}{4}-\frac{3 \bar{\sigma} _1 \bar{\sigma} _2}{2},\\
 \bar \nu_{\mathcal{R}1} & \simeq&-\bar{\epsilon} _1 \bar{\epsilon} _2-\frac{\bar{\epsilon} _2 \bar{\epsilon} _3}{2},
 \eqn
 and
 \bqn
 \bar \nu_{\mathcal{R}2} \equiv \frac{d^2 \nu_{\mathcal{R}}}{d\ln^2 (-\eta)} &=& \mathcal{O}(\bar \epsilon_i^3).
\eqn

Then, using the above expansions, the power spectrum for the curvature perturbation $\mathcal{R}$ can be calculated via Eq.(\ref{formula_pw}). After tedious calculations we obtain,
\begin{widetext}
\bqn\lb{PW_t}
\Delta_{\mathcal{R}}^2(k) &=&\bar A_s\Bigg[1-\left(2-2\bar D_p\right)\bar\epsilon _1 - \bar D_p\bar\epsilon _2+\bar\delta _1+\left(\bar D_a-\frac{9}{8}\right)\bar \zeta _1-\frac{3\bar \sigma _1}{4} +\left(\bar D_a-\frac{9}{8}\right)\bar \kappa _1+\left(\bar D_p^2-\bar D_p+\bar\Delta _1+2\bar \Delta _2+\frac{7 \pi^2}{12}-8\right)\bar\epsilon _1 \bar\epsilon _2\nb\\
&+&\left(\frac{\bar D_p^2}{2}+\frac{\bar\Delta _1}{4}+\frac{\pi ^2}{8}-\frac{3}{2}\right)\bar\epsilon _2^2+\left(-\frac{\bar D_p^2}{2}+\bar\Delta_2+\frac{\pi^2}{24}\right)\bar\epsilon_2\bar \epsilon _3+\left(2\bar D_p^2+2\bar D_p+\bar\Delta _1+\frac{\pi ^2}{2}-5\right)\bar\epsilon _1^2+\frac{9\bar \delta _1\bar \sigma _1}{4}\nb\\
   &+ &\left(-\frac{3\bar D_p}{4}-\frac{5}{2}\right)\bar\sigma _1\bar \sigma _2+\left(-2 \bar D_a \bar D_p-\frac{9031 \bar D_a}{2715}+\frac{9\bar  D_p}{4}+\frac{9}{4}\right)\bar\zeta _1\bar \epsilon _1-\frac{\bar\delta _1^2}{2}+\left(-2\bar D_p-2\right)\bar\delta _1 \bar\epsilon _1+\left(3 \bar D_a+\frac{9}{8}\right)\bar \delta _1 \bar \zeta _1\nb\\
   &+& \left(\bar D_a \bar D_p-\frac{449 \bar D_a}{724}-\frac{9\bar D_p}{8}-\frac{4405}{2896}\right)\bar\kappa _1 \bar\kappa _2+\frac{3}{4}\bar D_p\bar\sigma _1 \bar\epsilon _2-\frac{57 \bar\sigma _1^2}{32}+ \left(\bar D_p+2\right)\bar\delta _1\bar \delta _2- \bar D_p\bar\delta _1\bar \epsilon _2 +\left(3 \bar D_a-\frac{15}{8}\right)\bar \delta _1 \bar \kappa _1\nb\\
   &+& \left(-\bar D_a \bar D_p+\frac{1829\bar D_a}{5430}+\frac{9\bar  D_p}{8}-\frac{9}{8}\right)\bar\zeta _1 \bar\epsilon _2+ \left(-2 \bar D_a\bar  D_p-\frac{9031\bar  D_a}{2715}+\frac{9 \bar D_p}{4}+\frac{9}{4}\right)\bar\kappa _1 \bar\epsilon _1+\left(\frac{39}{32}-\frac{11 \bar D_a}{4}\right)\bar \kappa _1 \bar\sigma _1\nb\\
   &+&\left(-\bar D_a \bar D_p+\frac{1829\bar D_a}{5430}+\frac{9 \bar D_p}{8}-\frac{9}{8}\right)\bar\kappa _1 \bar\epsilon _2+\left(-\frac{11 \bar D_a}{4}-\frac{81}{32}\right)\bar\zeta _1 \bar\sigma _1+\left(\frac{297}{640}-\frac{69 \bar D_a}{10}\right)\bar\zeta _1^2+\left(\frac{3\bar  D_p}{2}+\frac{3}{2}\right)\bar\sigma _1\bar \epsilon _1\nb\\
   &-&\left(\frac{2457}{320}-\frac{79\bar D_a}{5}\right)\bar \zeta _1\bar\kappa _1+\left(\frac{3657}{640}-\frac{89 \bar D_a}{10}\right)\bar\kappa
   _1^2+ \left(\bar D_a \bar D_p-\frac{449 \bar D_a}{724}-\frac{9 \bar D_p}{8}-\frac{10197}{2896}\right)\bar\zeta _1 \bar\zeta _2\Bigg],
\eqn
\end{widetext}
where $\bar A_s\equiv\frac{181 \bar{H}^2}{72 e^3 \pi ^2 \bar{\epsilon} _1 M_{\text{pl}}}$, $\bar D_p\equiv\frac{67}{181}-\ln2$, $\bar D_a\equiv\frac{90}{181}$, $\bar\Delta _1\equiv\frac{183606}{32761}-\frac{\pi ^2}{2}$, and $\bar\Delta _2\equiv\frac{9269}{589698}$. Note that a letter with an over bar denotes quantity evaluated at the turning point $\bar y_0$. From the scalar spectrum (\ref{PW_t}), we observe that the effects of the four new operators affect the spectrum at the leading-order in the slow-roll expansion. We would like to mention that when we wrote down the action (\ref{pi_action}), we have neglected all the terms those do not contribute to the scalar power spectrum at the leading-order. But they may lead to contributions in the scalar spectrum at the second-order which are not included in the above expression. This is because their contributions only affect the spectral index $n_s$ and tensor-to-scalar ratio $r$ at the third-order and our intension here would be to obtain results which are accurate up to the second-order in the spectral index $n_s$ and $r$ for matching the accuracy of future observational data. 

%However, their contributions at the second-order are  not included   scalar spectral index $n_s$ and $r$ at the first two orders in the slow-roll expansion. This is because our intension here would be to obtain results which are accurate up to the second-order in the spectral index $n_s$ and tensor-to-scalar ratio $r$, as we mentioned in Sec. II. 
%
%These terms do not contribute to the scalar spectrum at the leading-order, but leads to contributions in power spectrum at the second-order. However, these contributions are not included in the second-order expressions of power spectrum derived in the above, because our intension here would be to obtain results which are accurate up to the second-order in the spectral index $n_s$ and $r$,  In this paper, we do not consider their However, they do contribute to the Considering  although we have derived the scalar power spectrum up to the second-order in the slow-roll expansion, 

Then with the scalar power spectrum given above, the scalar spectral index is,
\bqn
n_s-1&\simeq&-2 \bar\epsilon _1-\bar\epsilon _2-2\bar \epsilon _1^2+\bar\delta _1 \bar\delta _2- \left(2\bar D_n+3\right)\bar\epsilon _1 \bar\epsilon _2\nb\\
 &&-\bar D_n\bar\epsilon _2 \bar\epsilon _3-\frac{5 \bar\zeta _1 \bar\zeta _2}{8}-\frac{5 \bar\kappa _1 \bar\kappa _2}{8}-\frac{3\bar\sigma _1 \bar\sigma _2}{4},\nb\\
\eqn
and the running of the scalar spectral index is expressed as
\bqn
\alpha_s &\simeq&-2 \bar\epsilon_1 \bar\epsilon _2-\bar\epsilon _2 \bar\epsilon _3-6 \bar\epsilon _1^2 \bar\epsilon _2-\left(2\bar D_n+3\right)\bar\epsilon _1 \bar\epsilon _2^2 - \bar D_n\bar\epsilon _2 \bar\epsilon _3^2\nb\\
&&-\left(2\bar D_n+4\right)\bar\epsilon _1 \bar\epsilon _2 \bar\epsilon _3- \bar D_n\bar\epsilon _2 \bar\epsilon _3 \bar\epsilon _4+\bar\delta _1 \bar\delta _2^2+\bar\delta _1 \bar\delta _3 \bar\delta _2\nb\\
   &&-\frac{5}{8} \left(\bar\zeta _1 \bar\zeta _2^2+\bar \zeta _1 \bar\zeta _2\bar \zeta _3+\bar\kappa _1 \bar\kappa_2^2+\bar \kappa _1\bar \kappa _2 \bar\kappa _3\right)\nb\\
   &&-\frac{3}{4}\left(\bar \sigma _1\bar\sigma_2^2+ \bar\sigma _1 \bar\sigma _2 \bar\sigma _3\right).
\eqn
For scalar spectral index, the new effects denoted by the four new sets of the slow-roll parameters appear at the next-to-leading order, while for the running of the index, they only contribute to the third-order of the slow-roll approximation.

\subsection{Tensor Spectrum}

Now we consider the tensor spectrum. First we need to derive the expressions of $ \nu_{t 0}, \nu_{ t 1}, \nu_{t 2}$, and $c_{t 0}, c_{ t 1}, c_{ t 2}$. Repeating similar calculations for scalar perturbations, we obtain
\bqn
\bar \nu_{t 0} &=& \frac{3}{2}+\bar \epsilon _1+\bar \epsilon _1^2+\frac{4 \bar \epsilon _1 \bar \epsilon _2}{3}+\mathcal{O}(\bar \epsilon^3),~~~~~~~ \\
\bar \nu_{t1}&\equiv & \frac{d\nu_t}{d\ln(-\eta)} = -\bar \epsilon_1 \bar \epsilon_2 +\mathcal{O}(\bar \epsilon^3),\\
\bar \nu_{t 2} &\equiv & \frac{d^2\nu_t}{d\ln^2(-\eta)} =\mathcal{O}(\bar \epsilon^3),
\eqn
and
\bqn
\bar c_{t0}&=&1+\frac{\bar\kappa _1 \bar\epsilon _1}{2}+\mathcal{O}(\bar \epsilon_i^3),~~~~~~~~ \\
\bar c_{ t1} & \equiv & \frac{d c_{t}}{d\ln (-\eta)} = \mathcal{O}(\bar \epsilon_i^3).
\eqn

Then,  the power spectrum for the tensor perturbation $h_k$ reads
\bqn
\Delta_{t}^2(k) &=&\bar A_t\Bigg[1+ \left(-2 \bar D_p-2\right)\bar \epsilon _1\nb\\
&+&\left(2 \bar D_p^2+2\bar D_p+\bar \Delta _1+\frac{\pi ^2}{2}-5\right)\bar\epsilon _1^2 \nb\\
&+&\left(-\bar D_p^2-2\bar D_p+2\bar \Delta _2+\frac{\pi ^2}{12}-2\right)\bar \epsilon _1\bar \epsilon _2 -\frac{\bar \epsilon _1\bar \kappa_1}{2}\Bigg], \nb\\
\eqn
where $\bar A_t\equiv\frac{181 \bar H^2}{36 e^3 \pi ^2}$. Also the tensor spectral index and its running are given by
\bqn
n_t &\simeq&-2 \bar\epsilon _1-2\bar \epsilon _1^2-\left(\bar D_n+1\right)2 \bar\epsilon _1 \bar\epsilon _2,
\eqn
and
\bqn
\alpha_t\simeq&-&2 \bar\epsilon _1 \bar\epsilon _2-6 \bar\epsilon _1^2\bar\epsilon _2-\left(\bar D_n+1\right)2 \bar\epsilon _1 \bar\epsilon _2^2\nb\\
&-&2 \left(\bar D_n+1\right) \bar\epsilon _1 \bar\epsilon _2 \bar\epsilon _3.
\eqn
We observe that the new effects (represented by $ \bar \kappa_1$) can only affect the tensor spectrum at the next-to-leading order. In this case, the tensor spectral index and its running are keeping the same as the standard slow-roll inflation.

\subsection{Expressions at Horizon Crossing}

In the last two subsections, all the results are expressed in terms of quantities that are evaluated at the turning point. However, usually those expressions were expressed in terms of the slow-roll parameters which are evaluated at the time $\eta_\star$ when scalar or tensor perturbation modes cross the horizon, i.e., $a(\eta_\star) H (\eta_\star) = c_s(\eta_\star) k$ for scalar perturbations and $a(\eta_\star) H (\eta_\star) = c_t(\eta_\star) k$ for tensor perturbations. Consider modes with the same wave number $k$, it is easy to see that the scalar and tensor modes left the horizon at different times if $c_s(\eta) \neq c_t(\eta)$.  When $c_s(\eta_\star) > c_t(\eta_\star)$, the scalar mode leaves the horizon later than the tensor mode, and for $c_s(\eta_\star) < c_t(\eta_\star)$, the scalar mode leaves the horizon before the tensor one.

As we have two different horizon crossing times, it is reasonable to rewrite all our results in terms of quantities evaluated at the later time, i.e., we should evaluate all expressions at scalar horizon crossing time $a(\eta_\star) H (\eta_\star) = c_s(\eta_\star) k$ for
 $c_s(\eta_\star) > c_t(\eta_\star)$ and at tensor-mode horizon crossing $a(\eta_\star) H (\eta_\star) = c_t(\eta_\star) k$ for $c_s(\eta_\star) < c_t(\eta_\star)$.
\subsubsection{$c_s(\eta_\star) > c_t(\eta_\star)$}
Then, we shall re-write all the expressions in terms of quantities evaluated at the time when the scalar-mode horizon crossing $c_s(\eta_\star) > c_t(\eta_\star)$. Skipping all the tedious calculations, we find that scalar spectrum can be written in the form
\begin{widetext}
\bqn\lb{PW_h}
\Delta_{s}^2(k) &=& A^\star_s\Bigg[1-\left(2 D^\star_p+2\right)\epsilon_{\star1} -D^\star_p\epsilon _{\star2}-\left(D_a^\star-\frac{9}{8}\right)\zeta _{\star1}-\left(D_a^\star-\frac{9}{8}\right)\kappa_{\star1}+\delta _{\star1}-\frac{3 \sigma _{\star1}}{4}-\left(\frac{{D^\star_p}^2}{2}-\Delta^\star _2-\frac{\pi ^2}{24}\right)\epsilon _{\star2} \epsilon _{\star3}\nb\\
&+&\left(2 {D^\star_p}^2+2 D^\star_p+\Delta^\star _1+\frac{\pi^2}{2}-5\right)\epsilon _{\star1}^2+\left(\frac{{D^\star_p}^2}{2}+\frac{\Delta^\star_1}{4}+\frac{\pi ^2}{8}-1\right)\epsilon _{\star2}^2+\left({D^\star_p}^2-D^\star_p+\Delta^\star _1+2 \Delta^\star _2+\frac{7 \pi ^2}{12}-7\right)\epsilon _{\star1} \epsilon _{\star2}\nb\\
&-&\left(3 D_a^\star-\frac{15}{8}\right) \delta _{\star1} \kappa _{\star1}+\left(\frac{3D^\star_p}{4}-\frac{1}{2}\right)\sigma _{\star1} \epsilon _{\star2}-\left(D_a^\star D_p^\star+\frac{53117 D_a^\star}{17376}-\frac{9D^\star_p}{8}\right)\zeta _{\star1} \epsilon _{\star2}-\frac{2509}{320}D_a^\star \zeta _{\star1} \sigma_{\star1} \nb\\
&+&\left(3 D_a^\star+\frac{9}{8}\right) \delta _{\star1}\zeta _{\star1} + \left(\frac{3 D^\star_p}{2}+\frac{1}{2}\right)\sigma _{\star1} \epsilon _{\star1}+\left(D_a^\star D_p^\star-\frac{223033 D_a^\star}{28960}-\frac{9D_p^\star}{8}\right)\zeta _{\star1} \zeta _{\star2}+\frac{9 \delta_{\star1} \sigma _{\star1}}{4}-\frac{38187 D_a^\star \zeta _{\star1}^2}{6400}\nb\\
&+&\left(D_a^\star D_p^\star-\frac{191789 D_a^\star}{25920}-\frac{9 D_p^\star}{8}\right)\kappa _{\star1} \kappa _{\star2} -\left(\frac{3D^\star_p}{4}+\frac{5}{2}\right)\sigma _{\star1} \sigma _{\star2}-\left(2 D_a^\star D_p^\star+\frac{46213D_a^\star}{43440}-\frac{9 D_p^\star}{4}\right)\zeta _{\star1} \epsilon _{\star1}-\frac{\delta _{\star1}^2}{2} \nb\\
&-& \left(2 D_a^\star D_p^\star+\frac{46213 D_a^\star}{43440}-\frac{9 D_p^\star}{4}\right)\kappa _{\star1} \epsilon _{\star1}-\left(D_a^\star D_p^\star+\frac{46213 D_a^\star}{43440}-\frac{9 D_p^\star}{8}\right)\kappa _{\star1} \epsilon _{\star2}-\frac{57 \sigma _{\star1}^2}{32}\nb-\left(2 D^\star_p+2\right)\delta _{\star1} \epsilon _{\star1}\\
&-& D^\star_p\delta _{\star1} \epsilon _{\star2}-\frac{1147 D_a^\star \zeta _{\star1} \kappa _{\star1}}{3200}-\frac{287}{960} D_a^\star\kappa _{\star1}  \sigma _{\star1}+\frac{49759 D_a^\star \kappa _{\star1} ^2}{19200}+ \left(D^\star_p+2\right)\delta _{\star1} \delta _{\star2}\Bigg].
\eqn
\end{widetext}
where the subscript ``$\star$" denotes evaluation at the horizon crossing, $ A^\star_s\equiv\frac{181 {H_\star}^2}{72 e^3 \pi ^2 \epsilon _{\star1} }$, $ D^\star_p\equiv\frac{67}{181}-\ln3$, $ D_a^\star\equiv\frac{90}{181}$, $\Delta^\star _1\equiv\frac{485296}{98283}-\frac{\pi ^2}{2}$, and$\Delta^\star _2\equiv\frac{9269}{589698}$. For the scalar spectral index, one obtains
\bqn
n_s-1&\simeq&-2 \epsilon _{\star1}-\epsilon _{\star2}-2 \epsilon _{\star1}^2+\delta _{\star1} \delta _{\star2} \nb\\
&&-\left(2 D^\star_n+3\right)\epsilon _{\star1} \epsilon _{\star2}-D^\star_n\epsilon _{\star2} \epsilon _{\star3}\nb\\
&&-\frac{5 \zeta _{\star1} \zeta _{\star2}}{8}-\frac{5 \kappa _{\star1} \kappa _{\star2}}{8}-\frac{3\sigma _{\star1} \sigma _{\star2}}{4}.
\eqn
The running of the scalar spectral index reads
\bqn
\alpha_s&\simeq&-2 \epsilon_{\star1} \epsilon _{\star2}-\epsilon _{\star2} \epsilon _{\star3}-3 \epsilon _{\star1} \epsilon _{\star2}^2-6 \epsilon _{\star1}^2 \epsilon _{\star2}+\delta _{\star1} \delta _{\star2} \delta _{\star3}\nb\\
&&-4 \epsilon _{\star1} \epsilon _{\star2} \epsilon _{\star3}+\delta _{\star1} \delta _{\star2}^2-2 D^\star_n\epsilon _{\star1} \epsilon _{\star2}^2-\frac{3}{4} \sigma _{\star1} \sigma _{\star2} \sigma _{\star3} \nb\\
&&- D^\star_n\epsilon _{\star2} \epsilon _{\star3}^2-D^\star_n\epsilon _{\star2} \epsilon _{\star3} \epsilon _{\star4}-\frac{5}{8} \zeta _{\star1} \zeta _{\star2}^2-\frac{5}{8} \kappa _{\star1}\kappa _{\star2} \kappa _{\star3}\nb\\
&&-\frac{5}{8} \zeta _{\star1} \zeta _{\star2} \zeta _{\star3}-2 D^\star_n\epsilon _{\star1} \epsilon _{\star2} \epsilon _{\star3}-\frac{5}{8} \kappa _{\star1} \kappa _{\star2}^2-\frac{3}{4} \sigma _{\star1} \sigma _{\star2}^2.\nb\\
\eqn
Similar to the scalar perturbations, now let us turn to consider the tensor perturbations, which yield
\bqn
\Delta_{t}^2(k)&=&-2\left(1 + D^\star_p\right) \epsilon _{\star1}\nb\\
&&+\left(2 {D^\star_p}^2+2 D^\star_p-\Delta ^\star_1+\frac{\pi ^2}{2}-5\right)\epsilon _{\star1}^2\nb\\
&&+ \left(-{D^\star_p}^2-2 D^\star_p+\frac{\pi^2}{12}-2+2 \Delta^\star_2\right)\epsilon _{\star1} \epsilon _{\star2}\nb\\
&&-2 \delta _{\star1} \epsilon _{\star1}+3 \zeta _{\star1} \epsilon _{\star1}+\frac{\kappa _{\star1} \epsilon _{\star1}}{2}+\frac{3 \sigma _{\star1} \epsilon _{\star1}}{2}.\nb\\
\eqn
For the tensor spectral index, we find
\bqn
n_t\simeq-2 \epsilon _{\star1}-\left(2 D^\star_n+2\right)\epsilon _{\star1} \epsilon _{\star2}-2 \epsilon _{\star1}^2.
\eqn
Then, the running of the tensor spectral index reads
\bqn
\alpha_t\simeq&-&2 \epsilon _{\star1} \epsilon _{\star2}-6\epsilon _{\star1}^2\epsilon _{\star2} -2 \left(D^\star_n+1\right)\epsilon _{\star1}\epsilon _{\star2}^2\nb\\
&-&2\left(2 D^\star_n+1\right) \epsilon _{\star1} \epsilon _{\star2} \epsilon _{\star3}.
\eqn
Finally with both scalar and tensor spectra given above, we can evaluate the tensor-to-scalar ratio at the horizon crossing time$(\eta_\star)$, and find that
\begin{widetext}
\bqn
{r}&\simeq&16\epsilon _{\star1}\Bigg[1+D^\star_p\epsilon _{\star2}-\delta _{\star1}+\left(\frac{9}{8}-D^\star_a\right)\kappa _{\star1}+\left(\frac{9}{8}-D^\star_a\right)\zeta_{\star1}+\frac{3 \sigma _{\star1}}{4}+\left(D^\star_p-\Delta^\star _1-\frac{\pi ^2}{2}+5\right)\epsilon _{\star1} \epsilon _{\star2}\nb\\
&&+\left(\frac{{D^\star_p}^2}{2}-\Delta^\star_2-\frac{\pi ^2}{24}\right)\epsilon _{\star2} \epsilon _{\star3}+ \left(\frac{{D^\star_p}^2}{2}-\frac{\Delta^\star _1}{4}-\frac{\pi ^2}{8}+1\right)\epsilon _{\star2}^2 -\left(D^\star_p+2\right)\delta _{\star1}\delta _{\star2}+\left(\frac{3}{4} D^\star_p +\frac{1}{2}\right)\sigma _{\star1} \epsilon _{\star2}+\frac{421}{192}D^\star_a\kappa _{\star1} \sigma _{\star1}\nb\\
&&+\frac{3\delta _{\star1}^2}{2}+\left(\frac{9}{8}D^\star_p-D^\star_p D_a^\star+\frac{223033{D^\star_a}^2}{14400}\right) \zeta _{\star1} \zeta_{\star2}+ \left(\frac{9}{8}D^\star_p-D^\star_p D^\star_a+\frac{53117{D^\star_a}^2}{8640}\right) \zeta_{\star1} \epsilon _{\star2}-2 \delta _{\star1} \epsilon _{\star1}-\left(D^\star_a+\frac{27}{8}\right) \delta _{\star1} \zeta _{\star1}\nb\\
&&+\frac{623}{64}D^\star_a \zeta _{\star1} \sigma _{\star1}+\frac{869993 {D^\star_a}^2\zeta _{\star1}^2}{64000}-\frac{6252109{D^\star_a}^2\kappa_{\star1}^2}{1728000}+\left(\frac{9}{8}D^\star_p-D^\star_p D^\star_a+\frac{191789 {D^\star_a}^2}{25920}\right)\kappa _{\star1} \kappa_{\star2}+\frac{15481}{800}  {D^\star_a}^2 \zeta _{\star1} \epsilon _{\star1}\nb\\
&&+\frac{75 \sigma _{\star1}^2}{32}+\frac{1125697{D^\star_a}^2 \zeta _{\star1} \kappa_{\star1}}{288000}+\left(\frac{3}{4} D^\star_p +\frac{5}{2}\right)\sigma _{\star1} \sigma _{\star2}+\frac{5 \sigma _{\star1} \epsilon _{\star1}}{2}-\left(D^\star_a+\frac{3}{8}\right)\delta _{\star1} \kappa _{\star1}-\delta _{\star1} \epsilon _{\star2}-\left(D^\star_p-1\right)\delta _{\star1} \epsilon _{\star2}\nb\\
&&-\frac{15 \delta _{\star1} \sigma _{\star1}}{4}+\frac{598741{D^\star_a}^2\kappa _{\star1} \epsilon _{\star1}}{64800}+\left(\frac{9}{8}D^\star_p-D^\star_p D^\star_a+\frac{53117{D^\star_a}^2}{8640}\right)\kappa _{\star1} \epsilon _{\star2}\Bigg].
\eqn
\end{widetext}
\subsubsection{$c_s(\eta_\star) < c_t(\eta_\star)$}
For $c_s(\eta_\star) < c_t(\eta_\star)$, as the scalar mode leaves the horizon before the tensor mode does,we shall rewrite all the expressions in terms of quantities evaluated at the time when the tensor mode leaves the Hubble horizon $a(\eta_\star) H (\eta_\star) = c_t(\eta_\star) k$. Skipping all the tedious calculations, we find that the scalar spectrum can be written in the form
\begin{widetext}
\bqn\lb{PS}
\Delta_{s}^2(k) &=& A^\star_s\Bigg[1-\left(2 D^\star_p+2\right)\epsilon_{\star1} -D^\star_p\epsilon _{\star2}-\left(D_a^\star-\frac{9}{8}\right)\zeta _{\star1}-\left(D_a^\star-\frac{9}{8}\right)\kappa_{\star1}+\delta _{\star1}-\frac{3 \sigma _{\star1}}{4}-\left(\frac{{D^\star_p}^2}{2}-\Delta^\star _2-\frac{\pi ^2}{24}\right)\epsilon _{\star2} \epsilon _{\star3}\nb\\
&+&\left(2 {D^\star_p}^2+2 D^\star_p+\Delta^\star _1+\frac{\pi^2}{2}-5\right)\epsilon _{\star1}^2+\left(\frac{{D^\star_p}^2}{2}+\frac{\Delta^\star_1}{4}+\frac{\pi ^2}{8}-1\right)\epsilon _{\star2}^2+\left({D^\star_p}^2-D^\star_p+\Delta^\star _1+2 \Delta^\star _2+\frac{7 \pi ^2}{12}-7\right)\epsilon _{\star1} \epsilon _{\star2}\nb\\
&-&\left(3 D_a^\star-\frac{15}{8}\right) \delta _{\star1} \kappa _{\star1}+\left(\frac{3D^\star_p}{4}-\frac{5}{4}\right)\sigma _{\star1} \epsilon _{\star2}-\left(D_a^\star D_p^\star+\frac{175891 {D_a^\star}^2}{14400}-\frac{9D^\star_p}{8}\right)\zeta _{\star1} \epsilon _{\star2}-\frac{2509}{320}D_a^\star \zeta _{\star1} \sigma_{\star1} \nb\\
&+&\left(3 D_a^\star+\frac{9}{8}\right) \delta _{\star1}\zeta _{\star1} + \left(\frac{3 D^\star_p}{2}-1\right)\sigma _{\star1} \epsilon _{\star1}+\left(D_a^\star D_p^\star-\frac{223033 D_a^\star}{28960}-\frac{9D_p^\star}{8}\right)\zeta _{\star1} \zeta _{\star2}+\frac{9 \delta_{\star1} \sigma _{\star1}}{4}-\frac{38187 D_a^\star \zeta _{\star1}^2}{6400}\nb\\
&+&\left(D_a^\star D_p^\star-\frac{191789 D_a^\star}{25920}-\frac{9 D_p^\star}{8}\right)\kappa _{\star1} \kappa _{\star2} -\left(\frac{3D^\star_p}{4}+\frac{5}{2}\right)\sigma _{\star1} \sigma _{\star2}-\left(2 D_a^\star D_p^\star+\frac{102767{D_a^\star}^2}{7200}-\frac{9 D_p^\star}{4}\right)\zeta _{\star1} \epsilon _{\star1}-\frac{\delta _{\star1}^2}{2} \nb\\
&-& \left(2 D_a^\star D_p^\star+\frac{400727{ D_a^\star}^2}{64800}-\frac{9 D_p^\star}{4}\right)\kappa _{\star1} \epsilon _{\star1}-\left(D_a^\star D_p^\star+\frac{269683{D_a^\star}^2}{64800}-\frac{9 D_p^\star}{8}\right)\kappa _{\star1} \epsilon _{\star2}-\frac{57 \sigma _{\star1}^2}{32}-\left(D^\star_p-1\right)\delta _{\star1} \epsilon _{\star2}\nb\\
&-&2D^\star_p\delta _{\star1} \epsilon _{\star1}-\frac{1147 D_a^\star \zeta _{\star1} \kappa _{\star1}}{3200}-\frac{287}{960} D_a^\star\kappa _{\star1}  \sigma _{\star1}+\frac{49759 D_a^\star \kappa _{\star1} ^2}{19200}+ \left(D^\star_p+2\right)\delta _{\star1} \delta _{\star2}\Bigg].
\eqn
\end{widetext}
For the scalar spectral index, one obtains
\bqn
n_s-&1\simeq&-2 \epsilon _{\star1}-\epsilon _{\star2}-2 \epsilon _{\star1}^2+\delta _{\star1} \delta _{\star2}\nb\\
&&-\left(2 D^\star_n+3\right)\epsilon _{\star1} \epsilon _{\star2} - D^\star_n\epsilon _{\star2} \epsilon _{\star3}\nb\\
&&-\frac{5 \zeta _{\star1} \zeta _{\star2}}{8}-\frac{5 \kappa _{\star1} \kappa _{\star2}}{8}-\frac{3\sigma _{\star1} \sigma _{\star2}}{4}.
\eqn
The running of the scalar spectral index reads
\bqn
\alpha_s&\simeq&-2 \epsilon_{\star1} \epsilon _{\star2}-\epsilon _{\star2} \epsilon _{\star3}-3 \epsilon _{\star1} \epsilon _{\star2}^2-6 \epsilon _{\star1}^2 \epsilon _{\star2}+\delta _{\star1} \delta _{\star2} \delta _{\star3}\nb\\
&&-4 \epsilon _{\star1} \epsilon _{\star2} \epsilon _{\star3}+\delta _{\star1} \delta _{\star2}^2-2 D^\star_n\epsilon _{\star1} \epsilon _{\star2}^2-\frac{3}{4} \sigma _{\star1} \sigma _{\star2} \sigma _{\star3} \nb\\
&&- D^\star_n\epsilon _{\star2} \epsilon _{\star3}^2-D^\star_n\epsilon _{\star2} \epsilon _{\star3} \epsilon _{\star4}-\frac{5}{8} \zeta _{\star1} \zeta _{\star2}^2-\frac{5}{8} \kappa _{\star1}\kappa _{\star2} \kappa _{\star3}\nb\\
&&-\frac{5}{8} \zeta _{\star1} \zeta _{\star2} \zeta _{\star3}-2 D^\star_n\epsilon _{\star1} \epsilon _{\star2} \epsilon _{\star3}-\frac{5}{8} \kappa _{\star1} \kappa _{\star2}^2-\frac{3}{4} \sigma _{\star1} \sigma _{\star2}^2.\nb\\
\eqn
Similar to the scalar perturbations, now let us turn to consider the tensor perturbations, which yield
\bqn
\Delta_{t}^2(k)&=&-2\left(1 + D^\star_p\right) \epsilon _{\star1}\nb\\
&&+\left(2 {D^\star_p}^2+2 D^\star_p-\Delta ^\star_1+\frac{\pi ^2}{2}-5\right)\epsilon _{\star1}^2\nb\\
&&+ \left(-{D^\star_p}^2-2 D^\star_p+\frac{\pi^2}{12}-2+2 \Delta^\star_2\right)\epsilon _{\star1} \epsilon _{\star2}\nb\\
&&-2 \delta _{\star1} \epsilon _{\star1}+3 \zeta _{\star1} \epsilon _{\star1}+\frac{\kappa _{\star1} \epsilon _{\star1}}{2}+\frac{3 \sigma _{\star1} \epsilon _{\star1}}{2}.\nb\\
\eqn
For the tensor spectral index, we find
\bqn
n_t\simeq-2 \epsilon _{\star1}-\left(2 D^\star_n+2\right)\epsilon _{\star1} \epsilon _{\star2}-2 \epsilon _{\star1}^2.
\eqn
Then, the running of the tensor spectral index reads
\bqn
\alpha_t&\simeq&-2 \epsilon _{\star1} \epsilon _{\star2}-6\epsilon _{\star1}^2\epsilon _{\star2} -2 \left(D^\star_n+1\right)\epsilon _{\star1}\epsilon _{\star2}^2\nb\\
&&-2\left(2 D^\star_n+1\right) \epsilon _{\star1} \epsilon _{\star2} \epsilon _{\star3}.
\eqn
Finally with both scalar and tensor spectra given above, we can evaluate the tensor-to-scalar ratio at the horizon crossing time$(\eta_\star)$, and find that
\begin{widetext}
\bqn
{r}&\simeq&16\epsilon _{\star1}\Bigg[1+D^\star_p\epsilon _{\star2}-\delta _{\star1}+\left(\frac{9}{8}-D^\star_a\right)\kappa _{\star1}+\left(\frac{9}{8}-D^\star_a\right)\zeta_{\star1}+\frac{3 \sigma _{\star1}}{4}+\left(D^\star_p-\Delta^\star _1-\frac{\pi ^2}{2}+5\right)\epsilon _{\star1} \epsilon _{\star2}-D^\star_p\delta _{\star1} \epsilon _{\star2}\nb\\
&&+\left(\frac{{D^\star_p}^2}{2}-\Delta^\star_2-\frac{\pi ^2}{24}\right)\epsilon _{\star2} \epsilon _{\star3}+ \left(\frac{{D^\star_p}^2}{2}-\frac{\Delta^\star _1}{4}-\frac{\pi ^2}{8}+1\right)\epsilon _{\star2}^2 -\left(D^\star_p+2\right)\delta _{\star1}\delta _{\star2}+\left(\frac{3}{4} D^\star_p +\frac{5}{4}\right)\sigma _{\star1} \epsilon _{\star2}-\frac{15 \delta _{\star1} \sigma _{\star1}}{4}\nb\\
&&+\left(\frac{9}{8}D^\star_p-D^\star_p D_a^\star+\frac{223033{D^\star_a}^2}{14400}\right) \zeta _{\star1} \zeta_{\star2}-\left(D^\star_a+\frac{3}{8}\right)\delta _{\star1} \kappa _{\star1}+ \left(\frac{9}{8}D^\star_p-D^\star_p D^\star_a+\frac{175891{D^\star_a}^2}{14400}\right) \zeta_{\star1} \epsilon _{\star2}-2 \delta _{\star1} \epsilon _{\star1}\nb\\
&&+\frac{623}{64}D^\star_a \zeta _{\star1} \sigma _{\star1}+\frac{869993 {D^\star_a}^2\zeta _{\star1}^2}{64000}-\frac{6252109{D^\star_a}^2\kappa_{\star1}^2}{1728000}+\left(\frac{9}{8}D^\star_p-D^\star_p D^\star_a+\frac{191789 {D^\star_a}^2}{25920}\right)\kappa _{\star1} \kappa_{\star2}+\frac{15481}{800}  {D^\star_a}^2 \zeta _{\star1} \epsilon _{\star1}\nb\\
&&+\frac{75 \sigma _{\star1}^2}{32}+\frac{1125697{D^\star_a}^2 \zeta _{\star1} \kappa_{\star1}}{288000}+\left(\frac{3}{4} D^\star_p +\frac{5}{2}\right)\sigma _{\star1} \sigma _{\star2}+\frac{3\delta _{\star1}^2}{2}+\frac{598741{D^\star_a}^2\kappa _{\star1} \epsilon _{\star1}}{64800}+\frac{5 \sigma _{\star1} \epsilon _{\star1}}{2}-\left(D^\star_a+\frac{27}{8}\right) \delta _{\star1} \zeta _{\star1}\nb\\
&&+\left(\frac{9}{8}D^\star_p-D^\star_p D^\star_a+\frac{1058843{D^\star_a}^2}{129600}\right)\kappa _{\star1} \epsilon _{\star2}+\frac{421}{192}D^\star_a\kappa _{\star1} \sigma _{\star1}-\delta _{\star1} \epsilon _{\star2}\Bigg].
\eqn
\end{widetext}

\subsection{Comparison with results from generalized slow-roll approach}

In this subsection, we would like to provide a roughly comparison between our results and those derived by using the generalized slow-roll approach in the unified EFT of inflation \cite{Motohashi:2017gqb} at the leading slow-roll order. For this purpose, we can map the coefficients $C_{N}$, $C_{KK}$, $C_{NK}$, and $\tilde C_{KK}$ in Eq.~(16) in \cite{Motohashi:2017gqb} with the four new coefficients $M_2$, $\bar M_1$, $\bar M_2$, and $\bar M_3$ as
\bqn
C_{N}+\frac{1}{2}C_{NN}  &\to& - \dot H M_{\rm Pl}^2 + 2 M_2^4, \\
C_{NK} &\to& -  \bar M_1^3,\\
C_{KK} &\to& - \bar M_2^2 - M_{\rm Pl}^2,\\
\tilde C_{KK} &\to& - \bar M_3^2 + M_{\rm Pl}^2,\\
C_R &\to & \frac{1}{2}M_{\rm Pl}^2.
\eqn
In \cite{Motohashi:2017gqb}, the authors only consider the second-order spatial derivative terms in the action and make the fourth spatial derivative terms vanishing by imposing the condition
\bqn
C_{KK} = - \tilde C_{KK},
\eqn
which corresponds to $\bar M_2 ^2= - \bar M_3^2$ in the notation of this paper. One of essential differences of our calculation in this paper is that we have included the contributions of the fourth spatial derivative terms in the calculation of the power spectra. 

The scalar power spectrum in \cite{Motohashi:2017gqb} (Eq. (85) ) is calculated at the leading-order in the optimized slow-roll approximation. In order to compare with our results at least at the first-order in the slow-roll expansion, we have to recalculate their power spectra by including the $p=1$ order in Eq. (65) in \cite{Motohashi:2017gqb}. The scalar power spectrum can be calculated as (c.f. Eq.(65) in \cite{Motohashi:2017gqb})
\bqn
\ln\Delta_\zeta^2 \simeq G(\ln x_f) + p_1(\ln x_f) G'(\ln x_f),
\eqn
where
\bqn
G(\ln x_f) &\simeq& \ln\left(\frac{H^2}{8\pi^2b_s \epsilon_Hc_s}\right) - \frac{10}{3}\epsilon_H - \frac{2}{3}\delta_1 \nb\\
&&- \frac{7}{3}\sigma_{s1} - \frac{1}{3}\xi_{s1} - \frac{8}{3}\sigma_{s2},\\
G'(\ln x_f) &\simeq & 4 \epsilon_H + 2 \delta_1 + \sigma_{s1} +\xi_{s1} + \frac{2}{3}\delta_2 \nb\\
&&+ \frac{7 }{3}\sigma_{s2} + \frac{1}{3} \xi_{s2}, \\
p_1(\ln x_f) &=& \frac{7}{3}-\ln 2 - \gamma_E -  \ln x_f,
\eqn
with $\gamma_E$ being the Euler-Mascheroni constant. In the above expressions, $c_s$, $b_s$, $\delta_1$, $\epsilon_H$, $\sigma_{s1}$, $\sigma_{s2}$, $\xi_{s1}$, $\xi_{s2}$, $\delta_{s2}$, $x_1$, and $x_f$ are all defined in \cite{Motohashi:2017gqb}. In these quantities, $\sigma_{s1}$, $\sigma_{s2}$, $\delta_2$, $\xi_{s1}$, and $\xi_{s2}$ represent second-order contributions in the notation of our paper since they are defined as the derivatives of the slow-roll quantities $c_s$, $\delta_1$, $b_s$. Therefore we do not consider these terms when we compare the power spectrum at the leading-order. Then we can transform the above power spectrum in our notation, which is
\bqn
\Delta_\zeta^2 = \frac{H^2}{8 \pi^2 \epsilon_1} \left[1 +2 (\alpha_\star -1)\epsilon_1 +\alpha_\star\epsilon_2 - \frac{3}{4}\sigma_{1} +\delta_1 \right].\nb\\
\eqn
where $\alpha_\star \equiv 2- \ln 2 - \gamma_E \simeq 0.729637$ and the power spectrum is evaluated at the Horizon crossing point $x_f=1$. By comparing the coefficients of each term with those in (\ref{PS}) and considering the assumption $\zeta_1 = -\kappa_1$ if $\bar M_2^2= - \bar M_3^2$ in \cite{Motohashi:2017gqb}, we observe the two expressions are almost the same. The overall amplitude has a relative difference $\lesssim 0.15\%$, and the coefficients of $\epsilon_1$ and $\epsilon_2$ have relative differences 0.44\% and 0.16\%. The coefficients of $\sigma_1$ and $\delta_1$ are exactly the same in the two expressions. 

\section{Conclusion and Outlook}
\renewcommand{\theequation}{5.\arabic{equation}}\setcounter{equation}{0}
%%%%%%%%%%%%%%%%%%%%%%%%%%%%%%%
%%%%%%%%%%%%%%%%%%%%%%%%%%%%%%%}

The uniform asymptotic approximation method provides a powerful, systematically improvable, and error-controlled approach to construct accurate analytical solutions of linear perturbations. In this paper, by applying the third-order uniform asymptotic approximations, we have obtained explicitly the analytical expressions of power spectra, spectral indices, and running of spectral indices for both scalar and tensor perturbations in the EFT of inflation with the slow-roll approximation.

Comparing to the standard slow-roll inflation, the EFT of inflation introduces four new operators which can modify both the effective sound speed and the linear dispersion relation. In order to calculate the effects of these four new operators in the primordial spectra, we defined four new slow-roll parameters as in (\ref{fff}).  All the final expressions are  written in terms of both the Hubble flow parameters defined in (\ref{ff}) and the flow of four new slow-roll parameters and expanded up to the next-to-leading order in the slow-roll expansions so they represent the most accurate results obtained so far in the literature. We observe that the four new operators introduced in the action of the EFT of inflation can affect the primordial perturbation spectra at the leading-order and the corresponding spectral indices at the next-to-leading order. The running of the indices are keeping the same with that in the standard slow-roll inflation up to the next-to-leading order.

The next-to-leading order corrections to the scalar index $n_s$ and the tensor-to-scalar ratio $r$ in the slow- roll approximation are very important and useful in the future analysis with forthcoming observational data. As pointed out in \cite{abazajian_cmbs4_2016, huang_forecasting_2015}, the future CMB measurements can achieve the errors of the measurements on both $n_s$ and $r$ are $\lesssim 10^{-3}$. This implies, if the contributions of the new effects (for examples, $\delta_{\star 1} \delta_{\star 2}$, $\zeta_{\star 1} \zeta_{\star 2}$, etc) at the magnitude of $O(10^{-3})$, then they can not be ignored in the analysis with future experiments. Therefore, it would be interesting to constrain these new effects by using the more precise forthcoming observational data in the future. We expect that such constraints could help us to understand the physics of the early Universe.

\section*{Acknowledgements}
G.H.D., J.Q., Q.W., and T.Z. are supported by National Natural Science Foundation of China under the Grants No. 11675143, the Zhejiang Provincial Natural Science Foundation of China under Grant No. LY20A050002, and the Fundamental Research Funds for the Provincial Universities of Zhejiang in China under Grants No. RF-A2019015. A.W. is supported by National Natural Science Foundation of China with the Grants Nos. 11675145.
%11375153 (A.W.),
%, and 11105120 (T.Z.).

\appendix

\section{Integrals of $\sqrt{g}$ and the error control function}
\renewcommand{\theequation}{A.\arabic{equation}}\setcounter{equation}{0}

Here we present the calculation of the integral of $\sqrt{g}$ and the error control function $\mathscr{H}(+\infty)$ in the general expression of the power spectra (\ref{formula_pw}).

Let us first consider the integral of $\sqrt{g}$, which in general reads
\bqn\lb{intg}
\int_{y}^{\bar y_0} \sqrt{g(y')}dy' =\int_{y}^{\bar y_0} \sqrt{\frac{\nu^2(\eta)}{y'^2} - c^2(\eta) - b(\eta) y'^2} dy'.\nb\\
\eqn
Here we note that $\nu(\eta)$, $c(\eta)$, and $b(\eta)$ are all slow-varying quantities. In order to evaluate the integral with these time-dependent quantities, we consider expanding them about the turning point $y=\bar y_0$ as
\bqn
\lb{nu} \nu(\eta) = \bar \nu_0 + \bar \nu_1 \ln \frac{y}{\bar y_0} + \mathcal{O}\left(\ln^2\frac{y}{\bar y_0} \right),\\
\lb{c} c(\eta) = \bar c_0 + \bar c_1 \ln \frac{y}{\bar y_0} + \mathcal{O}\left(\ln^2\frac{y}{\bar y_0} \right),\\
\lb{b} b(\eta) = \bar b_0 + \bar b_1 \ln \frac{y}{\bar y_0} + \mathcal{O}\left(\ln^2\frac{y}{\bar y_0} \right),
\eqn
where
\bqn
\nu_1(\eta) \equiv \frac{d \nu(\eta)}{d \ln (- \eta)},\\
c_1(\eta) \equiv \frac{d c(\eta)}{d \ln (- \eta)},\\
b_1(\eta) \equiv \frac{d b(\eta)}{d \ln (- \eta)},
\eqn
and $\bar \nu_0 = \nu(\eta_0)$, $\bar \nu_1=\nu_1(\eta_0)$, $\bar c_0=c(\eta_0)$,  $\bar c_1=c_1(\eta_0)$,  $\bar b_0=b(\eta_0)$, and  $\bar b_1=b_1(\eta_0)$. With the above expansions, the integral (\ref{intg}) can be divided into two parts
\bqn\lb{intg}
\int_{y}^{\bar y_0} \sqrt{g(y')}dy' \simeq  I_0 +I_1,
\eqn
where
\bqn
I_0 &=& \int_{y}^{\bar y_0} \sqrt{\frac{\bar \nu_0^2}{y'^2} - \bar c_0^2 -\bar b_0 y'^2} dy,\\
I_1 &=&  - \int_{y}^{\bar y_0} \frac{\bar b_1 y'^4 + 2 \bar c_0 \bar c_1 y'^2 - 2 \bar \nu_0 \bar \nu_1}{2 y' \sqrt{\bar \nu_0^2 - \bar c_0^2 y'^2 - \bar b_0 y'^4}} \ln \frac{y'}{\bar y_0} dy'.\nb\\
\eqn
Performing the integral $I_0$ gives
\bqn\lb{I0}
\lim_{y \to 0} I_0 &=& -\frac{\bar \nu_0}{2} - \bar \nu_0 \ln \frac{y}{2 \bar \nu_0} -\frac{\bar \nu_0}{4}  \ln(\bar c_0^4 + 4 \bar b_0 \bar \nu_0^2)\nb\\
&-& \frac{\bar c_0^2}{4 \sqrt{\bar b_0}} \arctan\left( \frac{2 \sqrt{\bar b_0} \bar \nu_0}{\bar c_0^2}\right).
\eqn
For the second integral, since we only need to calculate the power spectrum up to the next-to-leading order in the slow-roll expansions, we can safely ignored the term with $\bar b_0$ in the squre root which only contributes to the third order corrections in the slow-roll expansions. With this simpfications, we have
\bqn
I_1 \simeq  - \int_{y}^{\bar \nu_0/\bar c_0} \frac{\bar b_1 y'^4 + 2 \bar c_0 \bar c_1 y'^2 - 2 \bar \nu_0 \bar \nu_1}{2 y' \sqrt{\bar \nu_0^2 - \bar c_0^2 y'^2}} \ln \frac{y'}{\bar \nu_0/\bar c_0} dy',\nb\\
\eqn
which leads to
\bqn\lb{I1}
\lim_{y \to 0} I_1 &\simeq& \frac{(1- \ln 2) \bar \nu_0}{\bar c_0}\bar c_1- \left(\frac{\pi^2}{24} - \frac{\ln^22}{2} + \frac{1}{2}\ln^2 \frac{y}{\bar \nu_0/\bar c_0}\right)\bar \nu_1\nb\\
&+& \frac{(5-6 \ln 2) \bar \nu_0^3 }{18 \bar c_0^4} \bar b_1.
\eqn

For error control function, the integral form is
\bqn
\mathscr{H}(\xi) &\simeq& \frac{5}{36} \left\{\int_{\bar y_0}^{y} \sqrt{g(y')} dy' \right\}^{-1}\Bigg|^{y}_{\bar y_0} \nb\\
&&- \int_{\bar y_0}^y \left\{\frac{q}{g} - \frac{5 g'^2}{16 g^3} + \frac{g''}{4 g^2}\right\} \sqrt{g(y')}dy'.\nb\\
\eqn
Using the expansions in Eqs.~(\ref{nu}, \ref{c}, \ref{b}), $\mathscr{H}$ can also divided into two parts,
\bqn
\mathscr{H}(\xi) \simeq \mathscr{H}_0(\xi) +\mathscr{H}_1(\xi), \lb{error}
\eqn
where
\bqn
 \lim_{y \to 0}\mathscr{H}_0(\xi) &\simeq& \frac{\bar c_0^4 + 8 \bar b_0 \bar \nu_0^2}{6 \bar c_0^4 \bar \nu_0 + 24 \bar b_0 \bar \nu_0^3}, \lb{H0}\\
 \lim_{y \to 0}\mathscr{H}_1(\xi)  &\simeq&  \frac{\bar c_1}{6 \bar c_0 \bar \nu_0} - \frac{23+12 \ln 2}{72 \bar \nu_0^2} \bar \nu_1 \nb\\
 &&+ \frac{(1+4 \ln 2) \bar \nu_0}{6 \bar c_0} \bar b_1. \lb{H1}
\eqn

%\bibliographystyle{ieeetr}
%\bibliography{Library1}

\end{document}